\documentclass[a4paper,12pt]{article}

\usepackage{epsfig}
\usepackage{lscape} % ENABLES environment landscape
\usepackage{amsmath,amsthm,amsfonts}
\usepackage{xcolor}
\usepackage{verbatim}
\newcommand {\E}      [1] {{\mbox{E}}\left[#1\right]}

\title{Covariate adjustment in randomised  trials: canonical link functions protect against model mis-specification}
\author{Ian R. White$^{1}$, Tim P Morris$^{1}$, Elizabeth Williamson$^2$ \vspace{2ex}\\
$^1$ \parbox[t]{\textwidth}{MRC Clinical Trials Unit at UCL,
Institute of Clinical Trials and Methodology,
90 High Holborn, London WC1V 6LJ, UK.}\vspace{2ex}\\
$^2$ Department of Medical Statistics, LSHTM, UK}

\begin{document}

% \renewcommand{\baselinestretch}{1}
% affects title/author page only

\maketitle

%%%%%%%%%%%%%%%%%%%%%%%%%%%%%%%%%%%%%%%%%%%%%%%%%%%%%%%%%%%%

\begin{abstract}
Covariate adjustment has the potential to increase power in the analysis of randomised trials, but mis-specification of the adjustment model could cause error. 
We explore what error is possible when the adjustment model omits a covariate by randomised treatment interaction, in a setting where the covariate is perfectly balanced between randomised treatments.
We use mathematical arguments and analyses of single hypothetical data sets.

We show that analysis by a generalised linear model with the canonical link function leads to no error under the null -- that is, if treatment effect is truly zero under the adjusted model then it is also zero under the unadjusted model. However, using non-canonical link functions does not give this property and leads to potentially important error under the null.
The error is present even in large samples and hence constitutes bias. 

We conclude that covariate adjustment analyses of randomised trials should avoid non-canonical links. 
If a marginal risk difference is the target of estimation then this should not be estimated using an identity link; alternative preferable methods include standardisation and inverse probability of treatment weighting.
\end{abstract}

%%%%%%%%%%%%%%%%%%%%%%%%%%%%%%%%%%%%%%%%%%%%%%%%%%%%%%%%%%%%

\section{Introduction}

Covariate adjustment is important in randomised controlled trials because it offers the promise of increasing the power to detect an effect of the randomised treatment (a treatment effect) \cite{Kahan++14}.
Covariate adjustment is usually done using a regression model including randomised treatment and a set of (preferably pre-specified) covariates. 
The regression model might be a linear model for quantitative data, a logistic regression for binary outcome data, or a linear mixed model or generalised linear mixed model for repeatedly measured data. 

It is important to define which treatment effect is to be estimated, and we call this the estimand \cite{ICHE9R1}.
Usually, the treatment effect is taken as the coefficient of randomised treatment in such models. 
In a linear model, covariate adjustment almost always leads to increased precision for this coefficient.
However, in non-linear models this is often not the case, and covariate adjustment instead increases the standard errors \cite{RobinsonJewell91}.
This paradox is explained by the fact that covariate adjustment changes the estimand \cite{Daniel++21};
when coefficients are compared on the same scale, covariate adjustment does increase their precision. 
Further, covariate adjustment increases power in all cases \cite{Neuhaus98}.

Sometimes, the treatment effect is instead derived from the fitted model. 
For example, the treatment effect on a binary outcome may be defined as the risk difference between experimental and control arms. 
This may be derived from the fitted model by standardisation, which predicts the outcome for each randomised individual's covariate pattern under each possible randomised treatment and takes averages over individuals \cite{Daniel++21}.
The treatment effect may also be derived by inverse probability of treatment weighting, which uses covariates to predict randomised treatment and then uses the fitted inverse probability of each individual having received their actual treatment as a set of weights that induces covariate balance in a subsequent unadjusted analysis \cite{ian:IPW_RCT}.

A common concern is that the more complex models involved in covariate adjustment are more open to model mis-specification and therefore to bias in the treatment effect \cite{RosenblumvanderLaan10}.
Two main forms of model mis-specification are possible.
Mis-specification of the main effects of the covariates, or omissions of covariate-covariate interactions, causes a failure to realise the full potential gain in power, but is unlikely to cause bias.
The omission of randomised treatment by covariate interactions, by contrast, has the potential to cause bias. 

We focus on analysis of randomised trials using generalised linear models (GLMs) \cite{McCullaghNelder89}. 
These regression models are defined by an outcome distribution and a link function. 
For each outcome distribution there is one link function, called the canonical link, which simplifies the form of the score equations. 
Most widely used GLMs have canonical link functions, e.g.\ linear regression (Normal distribution and identity link), logistic regression (binomial distribution and logit link) and Poisson  regression  (Poisson distribution and log link).
However non-canonical links are occasionally used, and in particular the model with a binomial distribution and identity link can be used to estimate a covariate-adjusted risk difference.

Other work has focussed on large-sample properties of covariate adjustment with mis-specified models. 
For example, even with mis-specified models, 
estimation of a marginal treatment effect using a generalised linear model (GLM) with canonical link is consistent \cite{Bartlett2018}; 
a wide range of hypothesis tests, based on GLMs with canonical links, asymptotically retain correct type 1 error \cite{RosenblumvanderLaan09};
and a standardisation approach using a canonical-link GLM within each trial arm is asymptotically unbiased and locally efficient \cite{RosenblumvanderLaan10}.
However, large-sample properties are of limited value when assessing the results of a single moderate-sized trial. 

This short note focuses on trials of any size and shows that a large degree of protection against model mis-specification is conferred by using generalised linear models with canonical link functions, and  conversely that using non-canonical link functions leads to potentially important error. 

%%%%%%%%%%%%%%%%%%%%%%%%%%%%%%%%%%%%%%%%%%%%%%%%%%%%%%%%%%%%

\section{Example} \label{sec:example}

% see file:
% N:\Home\Analysis\covariate_adjustment\examples with binary outcome\noncanonical link is dangerous in misspecified model.do

Table \ref{tab:hyp} shows hypothetical data from a two-arm randomised trial, where the binary outcome represents an unfavourable outcome. 
These data show that the experimental randomised treatment reduces the risk in the $X=0$ subgroup but increases the risk in the $X=1$ subgroup, in such a way that the overall effect of randomised treatment is zero.
That is, if this treatment were given to a population which (like the trial sample) consisted of equally sized $X=0$ and $X=1$ subgroups, then the overall change in risk would be zero.

In practice, analysis of these data might identify this interaction (it is just statistically significant, P=0.03).
However, if the interaction were unexpected and not specified in the statistical analysis plan, then analysis would probably proceed to estimate the overall treatment effect in all patients.

\begin{table}[ht]
\caption{Hypothetical data (number of events/number of individuals) from a two-arm randomised trial with binary outcome.}
\label{tab:hyp}
\centering
\begin{tabular}{lccc}\hline
Subgroup & $Z=1$ (Experimental)  &        $Z=0$ (Control)   & Risk difference (E-C)\\ \hline
       $X=0$ &      10/200 (5\%)  &       20/200 (10\%)  & -5\%\\
       $X=1$ &      90/200 (45\%) &       80/200 (40\%)   & 5\% \\\hline
    All &     100/400 (25\%) &      100/400 (25\%)   & 0\%\\\hline
\end{tabular}
\end{table}

We analysed these data using three link functions: logistic, identity and probit, and with and without adjustment for the covariate $X$. 
The estimand of interest is the treatment effect expressed as a risk difference.
We estimated this treatment effect as the coefficient of $X$ in the identity link model. In the models with other links,  we estimated the treatment effect using standardisation (using Stata's margins command \cite{Stata11}).
% introduced in Stata 11

The results in Table \ref{tab:hypres} show that, as expected, all unadjusted models report a treatment effect of zero with very similar standard errors. 
Among the adjusted analyses, only the logistic analysis reports a treatment effect of zero. 
Thus despite the model mis-specification, the logistic approach is correct.
This robustness is not seen for the non-canonical links, with the adjusted identity link model in particular wrongly estimating that randomised treatment reduces risk by 2.8\%.
These results are not a small-sample properties, since multiplying all counts in Table \ref{tab:hyp} by a large number would not change the point estimates. 
The error with a non-canonical link therefore represents both bias and inconsistency.

\begin{table}[ht]
\caption{Estimates of the marginal risk difference from the hypothetical data of Table  \ref{tab:hyp}, using GLMs with three link functions.}
\label{tab:hypres}
\centering
\begin{tabular}{lcccc}\hline
Link& \multicolumn{2}{c}{Unadjusted} &		\multicolumn{2}{c}{Adjusted} \\\cline{2-5}
&	Coefficient	& Standard 		&	Coefficient	& Standard 		\\ 
&		&  error		&		&  error		\\ \hline
Logit (canonical)&	0.000	&0.031	&0.000	&0.028\\
Identity (non-canonical)&	0.000	&0.031	&-0.028	&0.023\\
Probit (non-canonical)	&    0.000	&0.031	&-0.006	&0.028\\ \hline
\end{tabular}
\end{table}

%%%%%%%%%%%%%%%%%%%%%%%%%%%%%%%%%%%%%%%%%%%%%%%%%%%%%%%%%%%%

\section{Mathematical exploration}

This section shows that if (1) a GLM with canonical link is used and (2) the covariate $X$ is perfectly balanced across randomised treatments $Z$, then the coefficient of the $Z$ being zero in the unadjusted model implies that the coefficient of $Z$ is also zero in the unadjusted model. We assume a two-arm trial, but the argument generalises to more arms.

We write the unadjusted and adjusted analysis models as
\begin{eqnarray}
\E{Y|Z=z} &=& h(\alpha^* + \beta^* z) \label{eq:unadj:model} \\
\E{Y|Z=z,X=x} &=& h(\alpha + \beta z + \gamma x) \label{eq:adj:model}
\end{eqnarray}
where $h(.)$ is the inverse link function and $Y$ is the outcome.

The score equations for the unadjusted and adjusted models with canonical link are 
\begin{eqnarray}
\sum_i \left( \begin{array}{c}1\\z_i\end{array} \right) \{y_i - h(\alpha^* + \beta^* z_i)\} &=& 0\label{eq:unadj:score} \\
\sum_i \left( \begin{array}{c}1\\z_i\\x_i\end{array} \right) \{y_i - h(\alpha + \beta z_i + \gamma x_i)\} &=& 0 \label{eq:adj:score}
\end{eqnarray}
where $(x_i, z_i, y_i)$ are the values of $(X,Z,Y)$ for the $i$th individual, $i = 1 \ldots n$.
However, rather than estimating $\gamma$, we will show that the result holds for any given value of $\gamma$ (and hence that it holds for the value $\hat{\gamma}$). By fixing $\gamma$ we only need to solve the first two components of equation (\ref{eq:adj:score}). 

We assume that the unadjusted treatment effect is zero: that is, equation (\ref{eq:unadj:score}) has a solution with $\beta^*=0$.
This means that 
\begin{equation}\label{eq:score3}
\sum_i \left( \begin{array}{c}1\\z_i\end{array} \right) \{y_i - h(\alpha^*)\} = 0.
\end{equation}
We then show that whatever the value of $\gamma$ in model (\ref{eq:adj:model}), the value $\beta=0$ solves the first two components of equation  (\ref{eq:adj:score}).
The assumption of perfect balance implies that the average of $h(\alpha + \gamma x_i)$ in the $z_i=1$ subgroup is the same as its average across all individuals. Let this average be $\bar{h}(\alpha,\gamma)$. 
We can therefore write the adjusted model score as 
\begin{equation}\label{eq:scoreun}
\sum_i \left( \begin{array}{c}1\\z_i\end{array} \right) \{y_i - h(\alpha + \gamma x_i)\} 
=
\sum_i \left( \begin{array}{c}1\\z_i\end{array} \right) \{y_i - \bar{h}(\alpha,\gamma)\}.
\end{equation}
This shows that values $(\alpha^*, \beta^*)$ satisfying $h(\alpha^*)=\bar{h}(\alpha,\gamma)$ and $\beta^*=0$ solve equation (\ref{eq:unadj:score}). These values exist for any link function that is continuous and monotonic. Hence the adjusted treatment effect is also zero. 

The converse is also easy to show, that if the adjusted treatment effect is zero then the unadjusted treatment effect is also zero.

If instead a non-canonical link were used then the summands in equations (\ref{eq:unadj:score})  and (\ref{eq:adj:score}) would be multiplied by a further factor equal to the derivative of the canonical parameter with respect to the linear predictor, and this would depend on the covariate $x_i$. 
The above argument would therefore not hold. For example for a binary outcome with probability $\pi$, the canonical parameter is $\log\{\pi/(1-\pi)\}$. If logistic regression is used then the linear predictor is also $\log\{\pi/(1-\pi)\}$ and equations (\ref{eq:unadj:score})  and (\ref{eq:adj:score}) are correct. 
If a GLM with identity link is used then the linear predictor is $\pi$ and the adjusted model score equation becomes 
\begin{eqnarray}
\sum_i \left( \begin{array}{c}1\\z_i\\x_i\end{array} \right) \frac{y_i - h(\alpha + \beta z_i + \gamma x_i)}{h(\alpha + \beta z_i + \gamma x_i)(1-h(\alpha + \beta z_i + \gamma x_i))}   &=& 0. \label{eq:adj:score:noncan}
\end{eqnarray}

%%%%%%%%%%%%%%%%%%%%%%%%%%%%%%%%%%%%%%%%%%%%%%%%%%%%%%%%%%%%

\section{Systematic exploration}
% see file
% N:\Home\Analysis\covariate_adjustment\examples with binary outcome\adj_unadj_logit_binreg_v3.do

We explored data sets like that in Section \ref{sec:example}, with the same marginal distribution of $X$ and $Z$, but varying the numbers of positive outcomes in each cell from 10 to 20 in steps of 2.
This yielded $6^4$ data sets.
We analysed each data set using identity, log and logit links, unadjusted and adjusted for the covariate $X$.
To explore the impact of adjustment on the model coefficients, we did not perform the standardisation step in Section  \ref{sec:example}: coefficients are therefore not comparable between the different link functions. Unadjusted and adjusted coefficients are also not comparable for the logit link, though the latter difference tends to be small.

\begin{figure}
\caption{Bland-Altman plots comparing the unadjusted and adjusted estimates of the treatment effect over multiple data sets, for three different link functions.}
\label{fig:butterfly}
\centering
\includegraphics{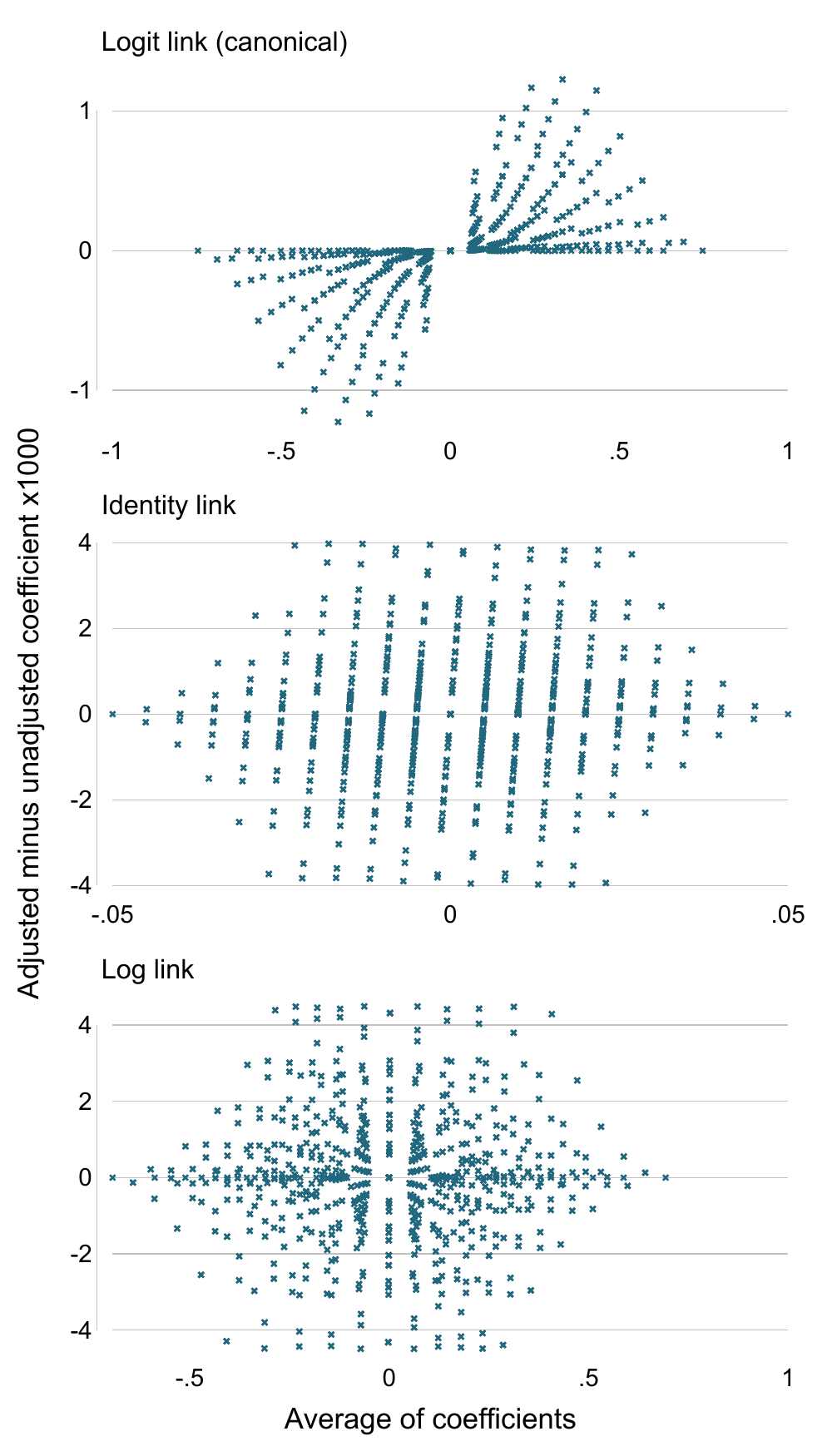}
\end{figure}

Figure \ref{fig:butterfly} shows Bland-Altman plots \cite{BlandAltman86} comparing the unadjusted and adjusted estimates of the treatment effect, for the three link functions.
The top panel, with the canonical link function, shows a pattern of the adjusted estimator being less than the unadjusted when the average is below zero, and more than the unadjusted when the average is above zero. This implies that the adjusted estimator is always more extreme (further from the null) than the unadjusted estimator. 
The middle and bottom panels, with non-canonical links, shows a very different and more diffuse pattern. In particular, there is a band of points with an average coefficient zero, implying that the unadjusted and adjusted coefficients may have different signs, even though the covariate is perfectly balanced across randomised treatments. 
This would be a worrying finding in the analysis of a randomised trial.
% i fact we don't get opposite signs, but we get plenty of ``unadj=0, adj ne 0''.

%%%%%%%%%%%%%%%%%%%%%%%%%%%%%%%%%%%%%%%%%%%%%%%%%%%%%%%%%%%%

\section{Discussion}

We have shown that canonical link functions provide special protection against error due to interactions when generalised linear models are fitted to randomised trial data.

It could be argued that the magnitudes of error that we have shown with non-canonical links -- for example, 3 percentage points on the risk difference -- are unimportant. The error would be larger if the interaction were larger. Here it could be argued that large interactions would always be detected.
However, there is an intermediate area where interactions are not statistically convincing and error remains of an important magnitude.
Further, in small data sets, even large interactions may be not statistically significant and hence may not be believed, and here error in covariate adjustment without a canonical link could be large.
We have not done a simulation study to demonstrate bias across many data sets, but nevertheless we have shown that these errors would arise in large samples and do constitute bias.

Fortunately there is a simple way to avoid bias, and that is to use canonical links. 
This means that for binary outcome data, estimation of the risk difference with covariate adjustment should not be done by directly fitting an identity-link binomial regression model, which has a non-canonical link.
A good alternative is to model the data using a canonical-link GLM and performing post-estimation standardisation, as we did in Section \ref{sec:example}.
Another alternative is to use inverse probability of treatment weighting in a simple canonical-link GLM. 
A separate paper more fully describes these approaches and their advantages and disadvantages \cite{arXiv:2107.06398}.

We have focussed on the case of perfect covariate balance. This has allowed us to focus on evaluating adjusted estimates when the unadjusted estimate is zero. 
Perfect covariate balance can realistically be achieved in trials with a small number of categorical covariates, provided the sample size is an exact multiple of the randomised block length, but it cannot be achieved in trials with continuous covariates. 
Our findings about the superiority of canonical links are likely to hold in trials that do not have perfect covariate balance, and this could be explored using simulation. 
However, some bias may still occur with canonical links. In other unpublished work, we have observed that covariate adjustment with a severely mis-specified model and a continuous covariate can led to small-sample bias, though not to large-sample bias. 
% see folder: N:\Home\Analysis\covariate_adjustment\ANCOVAbias

The implications of this work for analysis of time-to-event outcomes are unclear, since such outcomes are usually analysed by Cox models which are not in the GLM family. It could be that the close link of Cox models with canonical-link Poisson regression provides reassurance of robustness against model mis-specification.

%%%%%%%%%%%%%%%%%%%%%%%%%%%%%%%%%%%%%%%%%%%%%%%%%%%%%%%%%%%%

\section{Conclusions}

We conclude that non-canonical links should be avoided for covariate adjustment in randomised trials.
If a marginal risk difference is the target of estimation, then this should be estimated using logistic regression with standardisation or inverse probability of treatment weighting.

%%%%%%%%%%%%%%%%%%%%%%%%%%%%%%%%%%%%%%%%%%

\section*{Acknowledgements}

IRW and TPM were supported by the Medical Research Council Unit Programme number MC\_UU\_12023/21. 
EJW was supported by the MRC Network of Hubs for Trials Methodology HTMR Award MR/L004933/2/N96 and MRC project grant MR/S01442X/1.
We thank Brennan Kahan and Tim Clark for the encouragement to write this work up and Kelly van Lancker for helpful advice.

%%%%%%%%%%%%%%%%%%%%%%%%%%%%%%%%%%%%%%%%%%

%\bibliographystyle{unsrt}
%\bibliographystyle{c:/latex/bst/sim} % my Stat Med style
%\bibliographystyle{c:/latex/bst/jrss} %- requires natbib
\bibliography{c:/latex/library}

%%%%%%%%%%%%%%%%%%%%%%%%%%%%%%%%%%%%%%%%%%

%\clearpage
%
%% PLACE FIGURES AT TOP OF PAGES
%\makeatletter
%\setlength{\@fptop}{5pt}
%\makeatother
%
%\input{figures}
%
%%%%%%%%%%%%%%%%%%%%%%%%%%%%%%%%%%%%%%%%%%%
%
%\clearpage
%\input{tables}
%
%%%%%%%%%%%%%%%%%%%%%%%%%%%%%%%%%%%%%%%%%%%
%
%\clearpage
%\appendix
%\setcounter{page}{1}
%\setcounter{figure}{0}
%\renewcommand{\thefigure}{S\arabic{figure}}
%\renewcommand{\thepage}{S\arabic{page}}
%\section*{Supplementary Materials}
%
%\input{appendix}
%
%\clearpage
%
%
%\input{suppfigures}

%%%%%%%%%%%%%%%%%%%%%%%%%%%%%%%%%%%%%%%%%%%%%%%%%%%%%%%%%%%%%%%%%%%%%%%%%%%%%%%%

% tables
%\begin{landscape}
%\begin{table}[ht]
%\caption{CAPTION \label{tab:}}
%\begin{center}
%\begin{tabular}{lcccc}\hline
%CONTENTS
%\hline
%\end{tabular}
%\end{center}
%\end{table}
%\end{landscape}

%%%%%%%%%%%%%%%%%%%%%%%%%%%%%%%%%%%%%%%%%%%%%%%%%%%%%%%%%%%%%%%%%%%%%%%%%%%%%%%%

\end{document}